\documentclass[10pt, conference]{IEEEtran}
\usepackage{cite}
\usepackage{amsmath,amssymb,amsfonts}
\usepackage{algorithmic}
\usepackage{graphicx}
\usepackage{textcomp}
\usepackage{xcolor}
\usepackage{svg}
\usepackage{threeparttable}
\usepackage{siunitx}
\usepackage{booktabs}
\usepackage{subcaption}
\usepackage[draft]{hyperref}
\usepackage{geometry}
\def\BibTeX{{\rm B\kern-.05em{\sc i\kern-.025em b}\kern-.08em
    T\kern-.1667em\lower.7ex\hbox{E}\kern-.125emX}}
\begin{document}

\title{Learned Elevation Models as a Lightweight Alternative to LiDAR for Radio Environment Map Estimation}

\author{
\IEEEauthorblockN{Ljupcho Milosheski*, Fedja Močnik, Mihael Mohorčič, Carolina Fortuna}
\IEEEauthorblockA{\textit{Department of Communication Systems, Jožef Stefan Institute, Ljubljana, Slovenia} \\
\ *ljupcho.milosheski@ijs.si}
}

\maketitle

\begin{abstract}

Next-generation wireless systems such as 6G operate at higher frequency bands, making signal propagation highly sensitive to environmental factors such as buildings and vegetation. Accurate Radio Environment Map (REM) estimation is therefore increasingly important for effective network planning and operation. Existing methods, from ray-tracing simulators to deep learning generative models, achieve promising results but require detailed 3D environment data such as LiDAR-derived point clouds, which are costly to acquire, several gigabytes per \(\text{km}^2\) in size, and quickly outdated in dynamic environments. We propose a two-stage framework that eliminates the need for 3D data at inference time: in the first stage, a learned estimator predicts elevation maps directly from satellite RGB imagery, which are then fed alongside antenna parameters into the REM estimator in the second stage. Across existing CNN-based REM estimation architectures, the proposed approach improves RMSE by up to 7.8\% over image-only baselines, while operating on the same input feature space and requiring no 3D data during inference, offering a practical alternative for scalable radio environment modelling.

\end{abstract}


\section{Introduction}
\label{sec:intro}

Emerging applications such as Extended Reality (XR), Holographic Communication, and Digital Twins impose increasingly stringent requirements on wireless networks in terms of throughput and latency \cite{akyildiz2022wireless}. To meet these demands, next-generation systems such as 6G are moving towards higher frequency bands \cite{rohde2023xr}. However, higher frequency electromagnetic waves are inherently more sensitive to the environment: they experience greater attenuation, reduced diffraction around obstacles, and more significant blockage from buildings and vegetation. This necessitates far more careful and precise planning of network coverage compared to previous generations.

Radio Environment Map (REM) estimators are powerful tool for this purpose, enabling the prediction of signal propagation characteristics, such as pathloss, across a given area based on environment-specific inputs, including 3D geometry and antenna parameters. Approaches for REM estimation range from classical spatial interpolation methods \cite{pesko2014radio} and physics-based ray-tracing simulators such as Wireless InSite \cite{wireless_insite} and NVIDIA Sionna \cite{sionna}, to recently proposed Deep Learning (DL) generative models based on U-Net \cite{rml} and GAN-based architectures \cite{act_gan}. While these methods achieve remarkable results, they share a critical dependency on fine-grained 3D input data to accurately model the propagation environment, which limits their potential for large-scale deployments and dynamic environments. 

Acquiring such 3D data is highly demanding in practice, as LiDAR campaigns require specialized equipment, cover limited areas, and produce datasets of tens of gigabytes requiring intensive preprocessing~\cite{rml}. Open-source alternatives such as OpenStreetMap~\cite{openstreetmap} provide building footprints but omit vegetation, which can significantly affect signal propagation~\cite{Schampheleer2025}. Moreover, real-world environments change continuously due to construction and temporary structures, making 3D data-dependent estimators ill-suited for dynamic scenarios such as environment-aware network management and control \cite{zheng2021} of the next generation wireless networks. A promising alternative is to estimate scene geometry directly from aerial imagery using DL models~\cite{imele}, eliminating the need for dedicated LiDAR campaigns at inference time.

We propose a two-stage framework that eliminates the need for 3D data at inference time. A DL-based height estimator~\cite{imele} is trained on paired RGB and LiDAR data to predict elevation maps of the environment directly from aerial imagery, which is then used alongside antenna parameters as input to the REM estimator. Importantly, 3D data is required only during training, while at inference time, only RGB imagery is needed, which is cheaper to acquire in terms of storage and carbon footprint, and has the potential to be dynamically updated as the environment changes through on-demand platforms, such as \cite{vantor2026}. 

The contributions of this paper are as follows:
\begin{itemize}
    \item A two-stage REM estimation framework that replaces 3D data at inference time with a learned elevation estimator, achieving up to 7.8\% RMSE improvement over image-only baselines.
    \item A memory scalability analysis demonstrates that the proposed framework offers a practical alternative to 3D data-dependent methods for large-scale deployment.
    \item Model-agnostic validation of the REM estimation on three CNN-based architectures on the RMDirectionalBerlin benchmark, confirms the generalizability of the proposed approach.
\end{itemize}
This paper is organized as follows. Section \ref{sec:related} summaries related works, Sections \ref{sec:proposed} and \ref{sec:methodology} elaborate on the proposed framework and provide methodological details while Section \ref{sec:Results} analyzed the results. Section \ref{sec:conclusions} concludes the paper.

\section{Related Work}
\label{sec:related}

REM estimation has attracted growing attention in the wireless community in recent years, particularly with the emergence of Digital Twins, where REMs serve as a core component, capturing spatial distributions of radio propagation for network planning and optimization.

\textbf{Ray tracing and datasets.}
Advances in ray-tracing methods have progressed from primarily commercial packages~\cite{wireless_insite} to optimized open-source libraries such as Sionna~\cite{sionna}. 
Alongside these simulation tools, several high-quality datasets have been released to benchmark and train data-driven methods. OpenStreetMap~\cite{openstreetmap} provides 3D building models for major urban areas but lacks vegetation data. RadioMapSeer~\cite{RadioMapSeer} offers widely used simulated radio maps with estimated building heights. More detailed datasets have recently appeared, including the RMDirectionalBerlin dataset with paired LiDAR and aerial imagery~\cite{rml},
physics-informed radio maps~\cite{zhang2024physics}, and the multiband 3D SpectrumNet benchmark~\cite{SpectrumNet}.

\textbf{Deep learning approaches.}
Generative DL models have emerged as particularly promising for REM estimation due to their ability to capture complex propagation patterns. Wang et al.~\cite{RadioDiff} deploy a generative diffusion model to achieve accurate REM estimation from sparse measurements and environment geometry. Chen et al.~\cite{RemNet} employ dilated convolutional layers with an enlarged receptive field alongside a high-resolution feature-preserving architecture. GAN-based approaches such as ACT-GAN~\cite{act_gan} further demonstrate the effectiveness of generative methods. Collectively, these works highlight the strong potential of generative architectures for capturing the spatial complexity of radio propagation.

\textbf{Positioning of this work.}
To the best of our knowledge, only two published works~\cite{rml, wu2026} address REM estimation from aerial RGB imagery, both via end-to-end CNNs mapping RGB images and antenna parameters directly to path loss. In contrast, we decouple the problem into two stages, elevation estimation followed by REM prediction, hypothesizing that independent optimization of each task allows for greater specialization and stronger overall performance.

\section{Proposed approach}
\label{sec:proposed}
Let $I \in \mathbb{R}^{H \times W \times C}$ denote a satellite RGB image, $\mathbf{E} \in \mathbb{R}^{H \times W}$ a LiDAR-derived elevation map, $G \in \mathbb{R}^{d}$ a vector of transmitter metadata, including 3D coordinates, orientation angles, and antenna pattern identifiers, which are spatially projected as input features, and $R \in \mathbb{R}^{H \times W}$ a radio environment map (REM).
Conventional learning-based REM estimation models compute:
\begin{equation}
    R = f_\theta(\mathbf{E}, I, G)
    \label{eq:baseline}
\end{equation}

We propose to replace the physical elevation map with a learned estimator:
\begin{equation}
    \hat{\mathbf{E}} = g_\phi(I)
    \label{eq:elev}
\end{equation}
where $g_\phi$ is a neural image-to-elevation model trained on paired RGB and
3D data. The final REM prediction then becomes:
\begin{equation}
    \hat{R} = f_\theta(\hat{\mathbf{E}}, I, G)
    \label{eq:proposed}
\end{equation}
Training proceeds in two independent stages. First, $g_\phi$ is optimized to
minimize the elevation estimation error:
\begin{equation}
    \min_{\phi} \; \mathcal{L}_{\mathrm{elev}}(g_\phi(I), \mathbf{E})
\end{equation}
With $\phi$ fixed, $f_\theta$ is then trained to minimize the REM prediction error:
\begin{equation}
    \min_{\theta} \; \mathcal{L}_{\mathrm{REM}}(f_\theta(g_\phi(I), I, G), R)
\end{equation}

\begin{figure}[!t]
    \centering
    \includegraphics[width=\columnwidth]{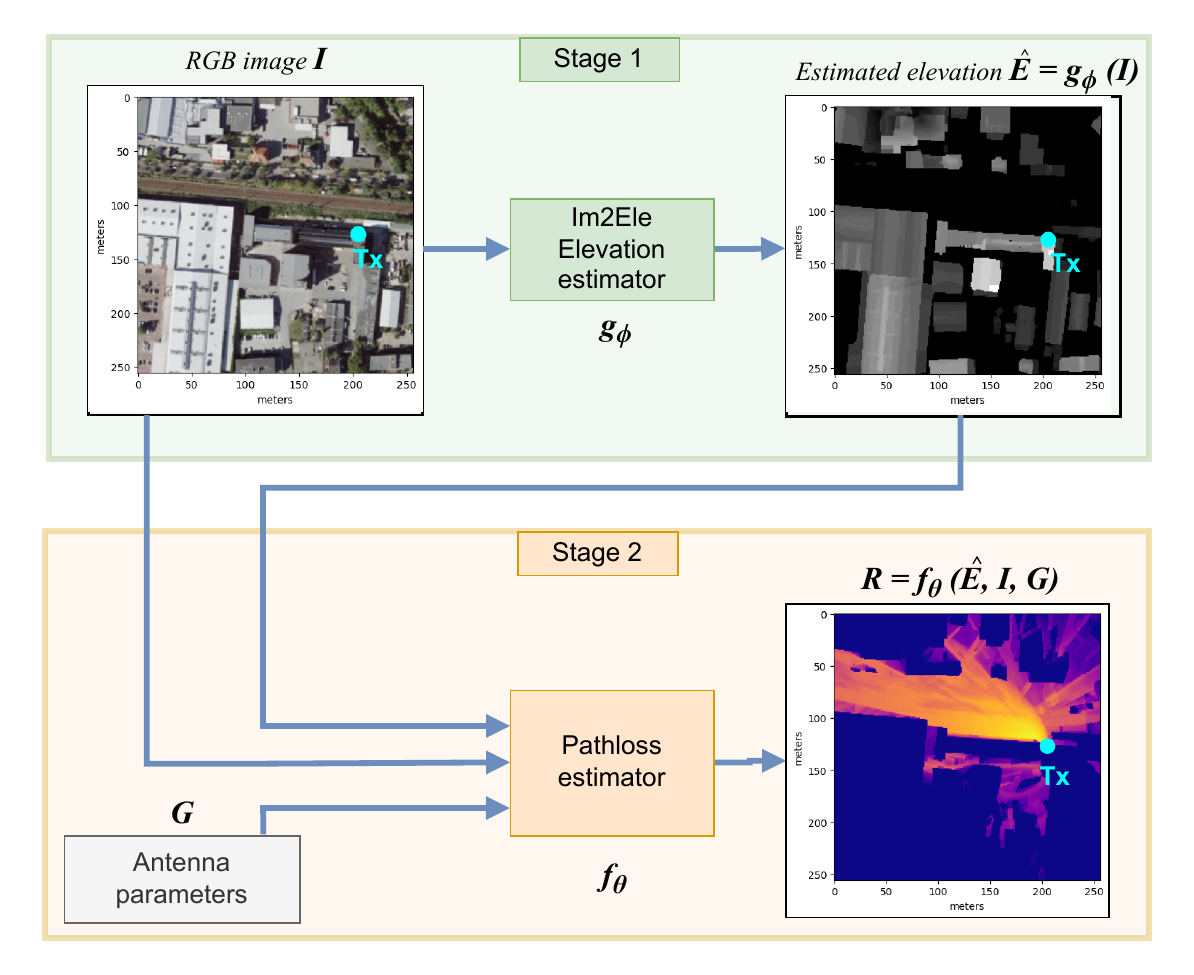}
    \caption{System diagram of the two-stage framework. Data and pathloss estimators from \cite{rml}, elevation estimator Im2Ele model from \cite{imele}.}
    \label{fig:sys_diagram}
\end{figure}

To realize $\hat{R}$ we propose a modular two-stage framework  illustrated in Fig.~\ref{fig:sys_diagram} as follows.

\subsubsection{Stage 1 — Elevation Estimation}

In the first stage, to obtain  $\hat{\mathbf{E}}$ from Eq.~\ref{eq:elev}, we retrain the Im2Ele model~\cite{imele} to learn $g_\phi(I)$ as depicted in the upper part of Fig.~\ref{fig:sys_diagram}.  Im2Ele is a SENet-based architecture that takes a single aerial RGB image as input and predicts a normalized Digital Surface Model (nDSM) representing the height of above-ground structures. We focus specifically on buildings, as these constitute the primary obstacles influencing radio propagation in urban environments. This focus also reduces scene variability, which is particularly important given that the RMDirectionalBerlin dataset provides only 424 image-elevation pairs, considerably fewer than the several thousand samples used in the original Im2Ele training~\cite{imele}, making generalization to geometrically inconsistent elements unreliable.

The elevation model is trained independently, and its weights are frozen prior to Stage 2 training, ensuring that the REM estimation model receives consistent elevation inputs throughout its optimization.

\subsubsection{Stage 2 — Radio Environment Map Estimation}

In the second stage, depicted in the lower part of Fig.~\ref{fig:sys_diagram} the pathloss estimator $f_\theta$, follows the experimental setup of Jaensch et al.~\cite{rml}. The model receives as input the aerial RGB image $I$, the estimated elevation map $\hat{\mathbf{E}}$ produced by Stage 1, a one-hot encoded transmitter location map, and antenna gain parameters $G$. These inputs are concatenated as separate feature channels and passed through the network, which outputs a spatial pathloss map $\hat{R} \in \mathbb{R}^{H \times W}$ representing radio signal propagation across the environment.

\subsection{Reference implementation and baseline selection}
\label{sec:validation setups}

To validate the proposed framework from Fig. \ref{fig:sys_diagram}, we propose a reference implementation as follows. Stage 1 is realized with Im2Ele as physical elevation map estimator. Stage II is realized via the model agnosting pathloss estimator evaluated via three different models trained using the RMDirectionBerlin and three neural architectures: LitRadioUNet, LitPMNet and LitUNetDCN~\cite{rml}. We refer to this proposed instantiation as \textbf{Im2Ele-predicted$\rightarrow$nDSM+images} elevation map.

We consider the following two baseline for evaluation purposes. Both baselines also include evaluations with the LitRadioUNet, LitPMNet and LitUNetDCN architectures on the RMDirectionBerlin dataset. We refer to the first baselines as \textbf{LiDAR$\rightarrow$nDSM+images}. It uses LiDAR-derived elevation alongside RGB imagery and antenna parameters, representing the best achievable performance given the richest available input data \cite{rml}. The second baseline, \textbf{Image-only}, uses aerial RGB imagery and antenna parameters without any elevation input, providing an end-to-end single-stage baseline with the least input information. This configuration is equivalent to the prior single-stage approaches of \cite{rml, wu2026}, and serves as the direct comparator against which the benefit of the proposed two-stage decoupling is measured.

\section{Methodology}
\label{sec:methodology}
In this section, we describe the dataset, training protocol, evaluation strategy, and baseline configurations used to assess the proposed two-stage pipeline.

\subsection{Training and Evaluation Data}
\label{sec:data}

All experiments were conducted using the RMDirectionalBerlin dataset, which contains 74,515 radio map samples across 424 RGB aerial images paired with corresponding normalized Digital Surface Models (nDSM) and transmitter metadata, including antenna gain parameters and location. The data covers both urban and vegetated areas, enabling separation of structural and natural height components.

While vegetation is known to affect signal propagation \cite{Schampheleer2025}, preliminary experiments incorporating vegetation into the elevation estimator degraded REM prediction performance, which we attribute to the high geometric variability of vegetation combined with the limited dataset size of 424 samples. Restricting the estimator to buildings reduces input variability and improves generalization under this data constraint. We adopted the training-validation-test splits of 80\%-10\%-10\%, accordingly.

\subsection{Model Training}
\label{sec:training}

Training was performed independently for the elevation estimation model (Im2Ele) and the radio propagation models (LitRadioUNet, LitPMNet and LitUNetDCN), ensuring optimization stability and preventing gradient interference between geometric reconstruction and radio propagation losses.

\subsubsection{Elevation Estimation}

The Im2Ele model was retrained on the RMDirectionalBerlin dataset to predict nDSM maps from monocular aerial imagery, following the normalization and augmentation strategy described in \cite{imele}, with one modification: color jitter augmentation was removed as color perturbations increased prediction instability in scenes containing large industrial roofs with high reflectance. Removing color jitter improved robustness in such failure cases without degrading overall performance. Thus, augmentation includes linear normalization of height values to range $[0,1]$, random geometric scaling, and random $8$-symmetry transformations (rotations and flips).

\subsubsection{Pathloss Estimation}

The radio propagation model receives as input the aerial RGBI image, an optional elevation map (ground-truth nDSM or estimated $\hat{H}$, clipped to the maximum height), a one-hot encoded transmitter location, and antenna gain parameters as per Eqs. \ref{eq:baseline} and \ref{eq:proposed}. All configurations were trained with Adam optimizer at a learning rate of $10^{-4}$, batch size 32, and FP16-mixed precision.

\subsection{Evaluation}
\label{sec:evaluation}


\subsubsection{Elevation and Pathloss}

Both stages of the pipeline are regression problems and are thus evaluated using the same standard metrics: the Mean Absolute Error is $\mathrm{MAE} = \frac{1}{n}\sum_{i=1}^{n} |y_i - \hat{y}_i|$ and the Root Mean Square Error is $\mathrm{RMSE} = \sqrt{\frac{1}{n}\sum_{i=1}^{n} (y_i - \hat{y}_i)^2}$.

For elevation estimation, $y_i$ and $\hat{y}_i$ denote ground-truth and predicted height values, and metrics are reported in meters. For radio map estimation, the same expressions apply with $y_i$ representing path loss values, scaled by the normalization factor $\mathrm{PL}_{\text{scale}}$ to express results in dB.
Beyond aggregate metrics, we also perform an error distribution analysis that examines the per-sample RMSE distribution across configurations to assess systematic shifts in prediction quality and the frequency of severe prediction failures.

\subsubsection{Memory Scalability}
To quantify the practical storage implications of replacing LiDAR with a learned elevation model, we compare the memory footprint of each approach related to the geographic coverage. We distinguish between the transient memory demands of the acquisition and preprocessing pipeline, which requires access to the full LiDAR point cloud regardless of how compact the final derived artefacts are. Since the transient LiDAR cost grows with the area covered while the learned model maintains a fixed deployment footprint regardless of scale, we report theoretical and empirical storage figures for both approaches on the RMDirectionalBerlin dataset, alongside peak training and inference memory consumption.

\subsubsection{Energy and Carbon Footprint}

To quantify the environmental impact of each pipeline, we analyze the carbon footprint of the proposed framework relative to the Image-based and 3D-dependent baseline, using commercially available drone platforms as a reference for data acquisition costs. Energy consumption of the Im2Ele model was measured using the eCAL library~\cite{eCAL}. Carbon emissions are estimated using the German grid carbon intensity of $363\ \mathrm{gCO_2e/kWh}$~\cite{uba2025}, equivalent to $1.008 \times 10^{-4}\ \mathrm{gCO_2/J}$, reflecting the geographic context of the RMDirectionalBerlin dataset. For data acquisition costs, we use the DJI Matrice 350 RTK as a reference platform, equipped with either the Zenmuse L2 for LiDAR or the Zenmuse P1 for RGB imagery, allowing a direct comparison under identical flight conditions with only the sensor payload exchanged. The reported LiDAR figures cover flight energy only and therefore represent a conservative lower bound, as the energy cost of preprocessing the raw point cloud is not accounted for and would further increase the carbon footprint of the 3D-dependent pipeline.

\section{Experimental Results} \label{sec:Results}

Following the methodology described in Section~\ref{sec:methodology}, we present results as follows: (i) a quantitative validation of the elevation estimator on the RMDirectionalBerlin; (ii), model-agnostic evaluation of the across the three setups described in Section~\ref{sec:validation setups}, considering the three different pathloss estimators, LitRadioUNet, LitUNetDCN and LitPMNet; (iii) Memory and Scalability and (iv) Energy and Carbon foorprint analysis from practical deployment perspective. The reported results are fully reproducible via the code available at our repository \footnote{https://github.com/sensorlab/REM-estimate}. 

\subsection{Elevation Estimation Performance}

\begin{figure}[htbp]
    \centering
    \includegraphics[width=\columnwidth]{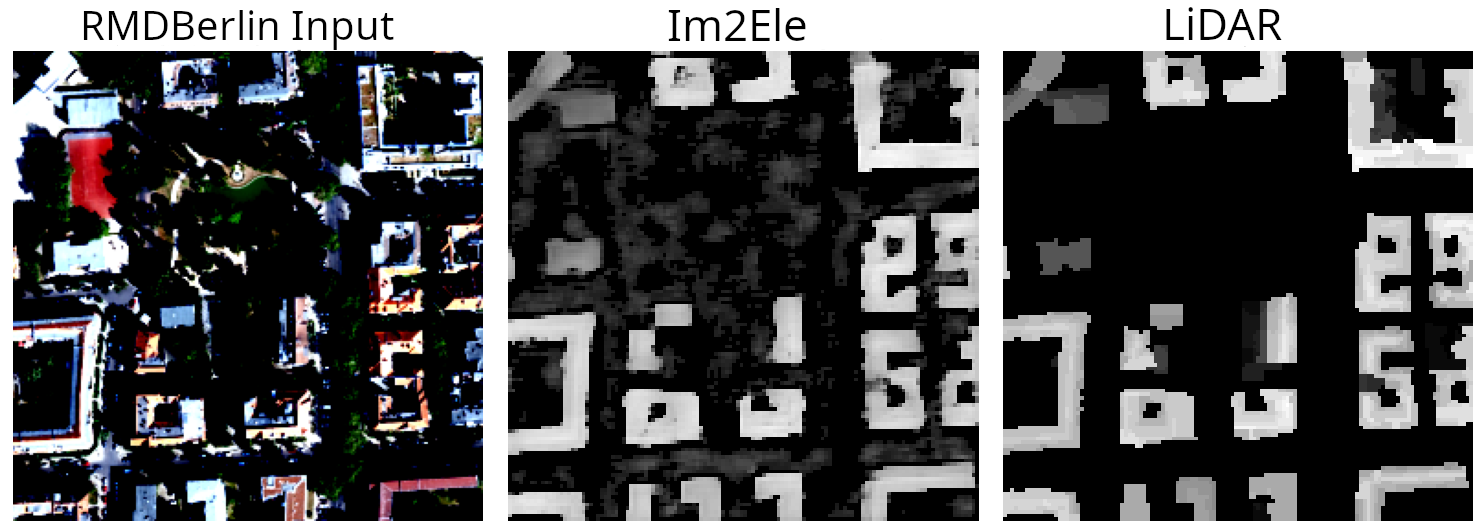}
    \caption{Qualitative examples comparing generated height maps with ground-truth LiDAR derived elevation maps.}
    \label{fig:gen-height-map}
\end{figure}

In Table \ref{tab:results}, the height estimation accuracy of the retrained Im2Ele model on the RMDirectionalBerlin dataset is compared with the originally reported results \cite{imele}. The first column outlines the configuration, the second specifies the dataset, and the third and fourth columns detail the MAE and RMSE metrics measured in meters. As can be seen from the table, the retrained model successfully minimizes average errors but struggles with localized extreme outliers compared to the baseline. Specifically, our retrained model achieves an MAE of 1.02 m, improving upon the baseline's 1.19 m. However, the RMSE increases slightly from 2.88 m in the baseline to 3.11 m in our retrained version. This is expected, considering that only buildings were used for ground-truth, as discussed in \ref{sec:data}, while in the RGB both vegetation and buildings are present. Figure \ref{fig:gen-height-map} presents an example of an input RGB image, with the corresponding generated height maps and ground-truth LiDAR-derived elevation maps. It can be observed that, while the generated maps reliably capture the general urban morphology, they also introduce height variation in the vegetated areas that exist in the RGB image.

\begin{table}[ht]
\label{tab:im2ele}
    \vspace{2mm}
    \centering
    \begin{threeparttable}
        \caption{Evaluation of Stage 1: comparison of performance of our retrained model with the results reported in Im2Ele~\cite{imele}.}
        
        \label{tab:results}
        
        \sisetup{
            detect-weight=true, 
            detect-inline-weight=math,
            group-digits=false 
        }

        \begin{tabular}{
            l 
            c                   
            c                   
            S[table-format=1.4]
            S[table-format=1.4] 
            S[table-format=1.4] 
            S[table-format=1.4] 
        }
        \toprule        
        \textbf{Configuration} & 
        \textbf{Dataset} & 
        {\textbf{MAE (m)}} & 
        {\textbf{RMSE (m)}} \\
        \midrule
        
        Im2Ele \cite{imele}    & IEEE DFC2018  & 1.19 & 2.88 \\
        Im2Ele        & RMDirectionalBerlin  & 1.02 & 3.11 \\
        
        \bottomrule
        \end{tabular}
    \end{threeparttable}
\end{table}

\subsection{Radio Map Estimation Performance}

Table \ref{tab:RMLresults} details the radio map prediction accuracy, presenting the RMSE for three different elevation input configurations: \textit{Image-only}, \textit{Im2Ele-predicted$\rightarrow$nDSM+image}, and \textit{LiDAR$\rightarrow$nDSM+image}. The results show that incorporating learned elevation information provides a consistent and measurable improvement over the image-only baseline. The proposed \textit{Im2Ele-predicted$\rightarrow$nDSM+image} configuration achieves a total Test RMSE of 0.0901 with the LitRadioUNet architecture, 0.0847 with the LitUNetDCN, and 0.0842 with LitPMNet, respectively, outperforming the \textit{Image-only} baseline by 4.3\%, 4.3\% and 7.8\%, respectively. A similar pattern is observed for the MAE values. Although the results still fall short of the LiDAR upper bound across the architectures, which reflects the inherent information loss from replacing physical elevation measurements with monocular image-based estimates, it is realistic to expect further narrowing of this gap if larger training datasets are utilized.

\begin{table}[ht] 
    \centering
        \caption{Evaluation of Stage 2: comparison of training configurations and results (RMSE, MAE). Our results are in bold.}
        \setlength{\tabcolsep}{4pt} 
        \label{tab:RMLresults}
        \footnotesize
        \sisetup{
            detect-weight=true,
            detect-inline-weight=math,
            group-digits=false
        }

        \begin{tabular}{
            l
            c
            c
            c
        }
            \toprule
            \textbf{Configuration} &
            \textbf{Architecture} &
            \textbf{RMSE} &
            \textbf{MAE}
            \\
            \midrule
            Image-only  & \text{LitRadioUNet} & $0.0942$ & $0.0417$ \\
            LiDAR$\to$nDSM+images  & \text{LitRadioUNet} & $0.0735$ & $0.0313$ \\
            Im2Ele-predicted$\to$nDSM+images  & \text{LitRadioUNet} & $\mathbf{0.0901}$ & $\mathbf{0.0401}$ \\
            \\
            Image-only  & \text{LitUNetDCN} & $0.0.0885$ & $0.0400$  \\
            LiDAR$\to$nDSM+images  & \text{LitUNetDCN} & $0.0684$ & 0.0304 \\
            Im2Ele-predicted$\to$nDSM+images  & \text{LitUNetDCN} & $\mathbf{0.0847}$ & $\mathbf{0.0380}$ \\
            \\
            Image-only  & \text{LitPMNet} & $0.0918$ & $0.0408$ \\
            LiDAR$\to$nDSM+images  & \text{LitPMNet} & $0.0686$ & $0.0295$ \\
            Im2Ele-predicted$\to$nDSM+images  & \text{LitPMNet} & $\mathbf{0.0856}$ & $\mathbf{0.0372}$ \\
            \bottomrule
        \end{tabular}
        
\label{tab:pl-eval}
\end{table}

\begin{figure*}[!t]
    \centering
    \begin{subfigure}{0.3\linewidth}
        \centering
        \includegraphics[width=\linewidth]{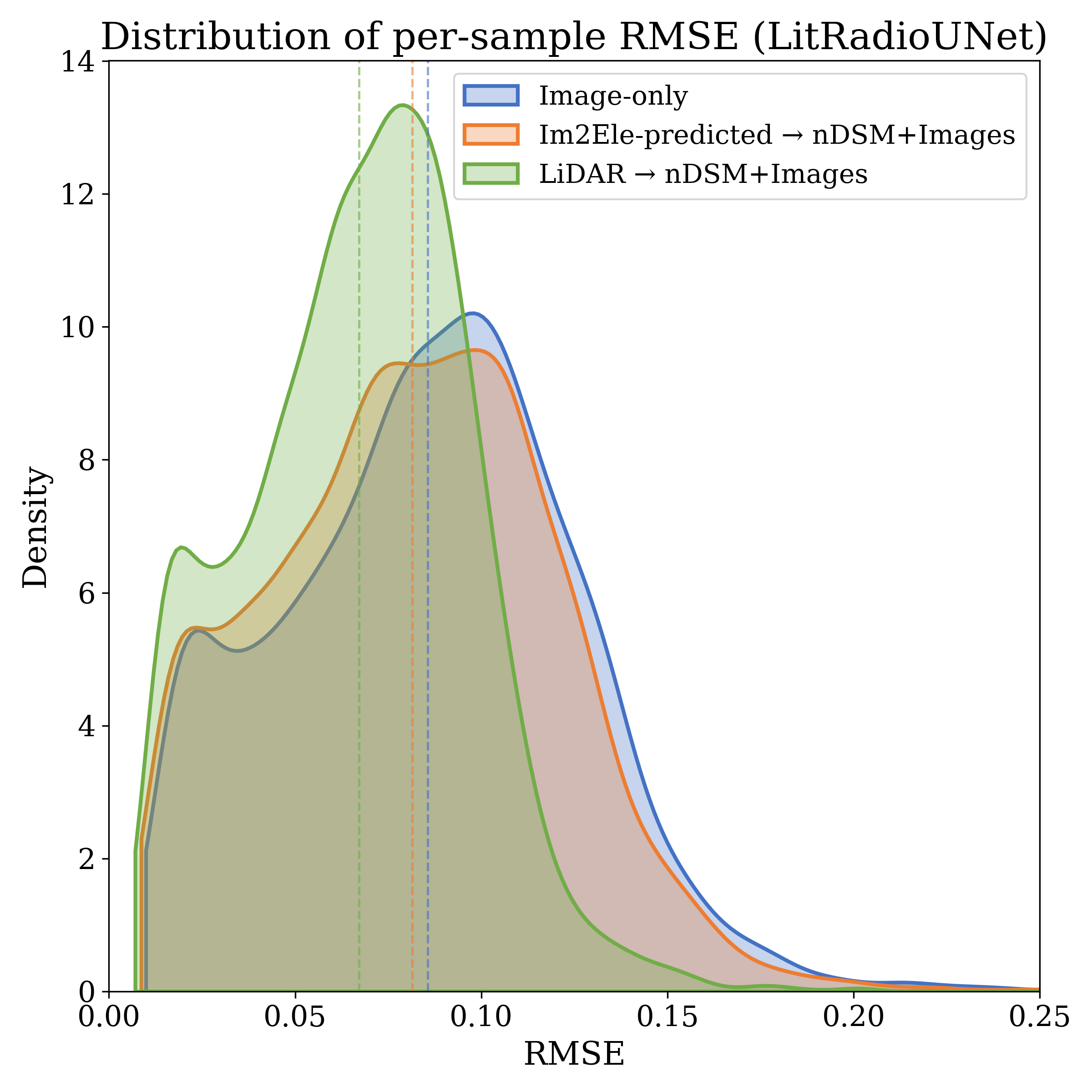}
        \caption{LitRadioUNet}
        \label{fig:err-dist-total}
    \end{subfigure}
    \begin{subfigure}{0.3\linewidth}
        \centering
        \includegraphics[width=\linewidth]{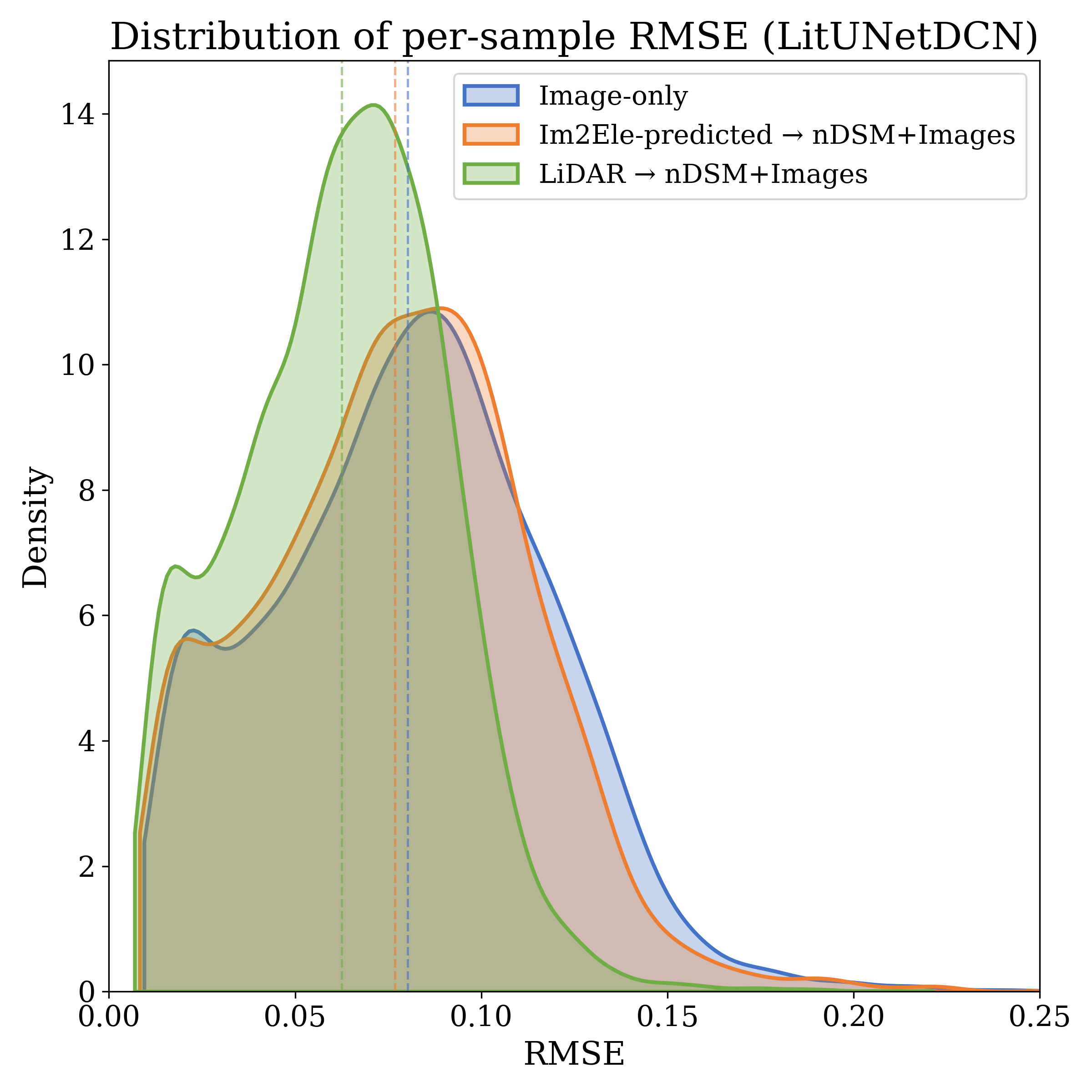}
        \caption{LitUNetDCN}
        \label{fig:err-dist-glf}
    \end{subfigure}
    \begin{subfigure}{0.3\linewidth}
        \centering
        \includegraphics[width=\linewidth]{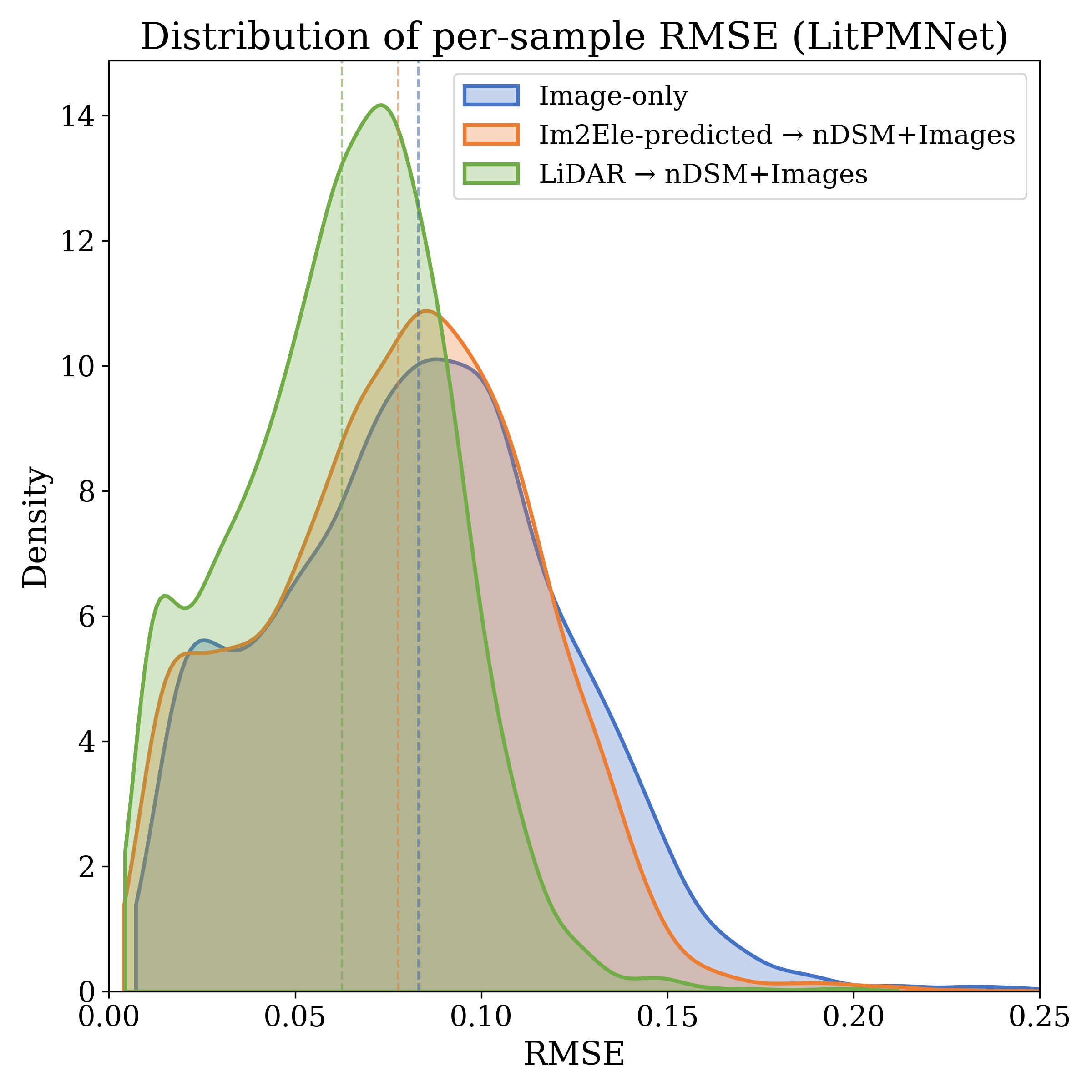}
        \caption{LitPMNet}
        \label{fig:err-dist-gf}
    \end{subfigure}
    \caption{Probability density functions of per-sample RMSE for the three configurations: image only, generated nDSM, and ground-truth nDSM.}
\end{figure*}

\subsubsection{Error Distribution Analysis}

The results of the error distribution study are presented in Figure~\ref{tab:RMLresults}, which plots the probability density function (PDF) of the per-sample RMSE across the three evaluated configurations, where the Image-only is colored blue, \textit{Im2Ele$\rightarrow$nDSM+Images} is red and \textit{LiDAR$\rightarrow$nDSM+Images} is green. We observe that the Im2Ele$\rightarrow$nDSM+image provides a systematic and measurable pathloss estimation accuracy improvement over the image-only baseline, successfully shifting the overall error distribution toward lower values and reducing the frequency of severe prediction failures. This is also evident from the vertical dashed lines representing the mean RMSE showing the shifts to the left relative to the Image-only baseline (blue), reducing the average error as reported in Table~\ref{tab:pl-eval}.

It can be observed that the peak density for the \textit{Im2Ele$\rightarrow$nDSM+image} occurs at a lower RMSE compared to the Image-only baseline. Furthermore, the \textit{Im2Ele$\rightarrow$nDSM+image} curve exhibits a lower density in the high-error region where RMSE exceeds 0.125. These observations indicate that while the learned elevation model does not reach the upper bound performance of the \textit{LiDAR$\rightarrow$nDSM+Images} (green), it effectively mitigates extreme prediction errors and provides a more reliable estimation than imagery alone.

\subsection{Memory and Scalability: Data vs.\ Model}

Beyond predictive performance, an important consideration is memory scalability. LiDAR-based approaches impose a large transient memory cost during acquisition and preprocessing that grows linearly with geographic coverage, although the final preprocessed artifacts are compact. In contrast, the learned elevation model has a fixed storage footprint regardless of the area covered.

We demonstrate this on the RMDirectionalBerlin dataset, which consists of 424 disjunct tiles of size $256\,\text{m} \times 256\,\text{m}$, covering a total area of $27.79\,\text{km}^2$. The raw LiDAR data is stored in LAS~1.4 Type~6 format at 30\,bytes/point and a point cloud density of 9.8\,points/m$^2$~\cite{rml}, corresponding to a theoretical uncompressed size of 8.17\,GB. After lossless compression into LAZ, the full dataset measures 249\,GB, equivalent to 294\,MB/km$^2$, and would exceed 1\,TB in raw LAS format. This dataset must be retained and processed during the preprocessing stage, including filtering and point cloud segmentation. The derived nDSM height maps produced, which occupy merely a few megabytes per region. This preprocessing stage therefore imposes a transient but unavoidable processing and memory demand that cannot be bypassed regardless of how compact the final artefacts are.

In contrast, Image-only and nDSM+image$\to$Im2Ele require no LiDAR data at inference time. The deployed Im2Ele model occupies 606\,MB, and the height reference input $\mathbf{H}$ at inference requires only 11\,MB, yielding a total deployment footprint that remains fixed as the number of covered regions grows. The peak memory consumption during training was 20\,GB (including framework and training overhead), while inference required approximately 600\,MB. This demonstrates the fundamental scalability difference: LiDAR storage and preprocessing impose a transient cost of $\mathcal{O}(A)$ that recurs for every new deployment region, whereas the neural representation maintains a constant footprint of $\mathcal{O}(1)$, yielding increasing memory and operational savings at scale.

\subsection{Energy and Carbon Footprint}

For LiDAR acquisition, the manufacturer-stated coverage of 2.5\,km$^2$ per flight at 150\,m requires 12 flights for the full 27.79\,km$^2$ dataset, consuming 22,741\,kJ (2,293\,g\,CO$_2$) based on the onboard battery capacity of 526.4\,Wh (2$\times$TB65). For RGB acquisition, the Zenmuse P1 (35\,mm lens) at its manufacturer-specified 240\,m operating altitude achieves a ground sampling distance of 3\,cm/px, nearly seven times finer than the 20\,cm/px imagery used as training input, with a coverage of 3\,km$^2$ per flight, requiring 10 flights and consuming 18,950\,kJ (1,911\,g\,CO$_2$). All figures cover flight energy only and represent lower bounds, excluding transportation, operator costs, and preprocessing compute, which is expected to be much more significant for the LiDAR data.

Table~\ref{tab:carbon} summarizes the estimated energy and carbon footprint of each approach. The one-time training of Im2Ele requires
411.6\,kJ (41.50\,g\,CO$_2$), after which each inference consumes only 0.16\,kJ (0.02\,g\,CO$_2$) per tile. For the full 424-tile
dataset, the total inference energy amounts to 67.84\,kJ and 6.84\,g\,CO$_2$. In the case of potential repeated surveys over the same region due to environmental changes, the RGB campaign is estimated to require approximately $1.2\times$ less energy than LiDAR.

Looking further ahead, the RGB modality opens a pathway toward fully eliminating ground-based acquisition. Commercial satellite imagery providers such as Vantor~\cite{vantor2026} offer on-demand high-resolution imagery at sub-50\,cm/px GSD, comparable to the training data. Adapting the proposed framework to satellite RGB inputs would remove the need for any physical survey for any deployment region worldwide.

\begin{table}[ht]
    \centering
    \caption{Estimated energy and carbon footprint of data acquisition
    and inference methods.}
    \label{tab:carbon}
    \begin{threeparttable}
    \begin{tabular}{l c c}
        \toprule
        \textbf{Method} &
        {\textbf{Energy (kJ)}} &
        {\textbf{CO\textsubscript{2} (g)}} \\
        \midrule
        Drone LiDAR survey     & 22741  & 2293 \\
        Drone RGB survey       & 18950  & 1911 \\
        \midrule
        Im2Ele training (one-time)      & 411.6  & 41.50 \\
        Im2Ele inference per tile       & 0.16   & 0.02  \\
        Im2Ele inference full dataset   & 67.84  & 6.84  \\
        LitRadioUNet inference per tile    & 0.004  & {$<$0.01} \\
        \bottomrule
    \end{tabular}
    \end{threeparttable}
\end{table}

\section{Conclusions and Future Work}
\label{sec:conclusions}

This paper presented a two-stage framework for Radio Environment Map estimation that eliminates the need for LiDAR data at inference time by replacing it with a learned elevation model. The first stage trains an image-to-elevation model (Im2Ele) on paired satellite imagery and LiDAR data, while the second stage uses the predicted elevation maps as input to a radio propagation model. Evaluated on the RMDirectionalBerlin dataset, the proposed approach improves RMSE by up to 7.8\% over the image-only baselines while using similar input features.

Beyond predictive accuracy, the framework offers substantial practical advantages large-scale REM modeling and digital twin applications. The learned elevation model requires only 606\,MB of storage regardless of geographic coverage, compared to LiDAR data whose storage grows linearly with area 294 MB per ${km}^2$ for the RMDirectionalBerlin dataset alone.

Future work will focus on validation across diverse geographic conditions, considering urban and rural areas, and on scaling the elevation estimator to larger training datasets, as well as considering satellite imagery as a deployment-phase data source.

\section*{Acknowledgments}
This work was supported in part by the Slovenian Research Agency (ARIS) under grant P2-0016.

\bibliographystyle{IEEEtran}
\bibliography{main}

\end{document}